\documentclass[9pt,twocolumn,twoside]{pnas-new}

\templatetype{pnasresearcharticle} 

\usepackage{amsmath}
\usepackage{amssymb}
\usepackage{bm}

\newcommand{\revise}{}
\newcommand{\change}{}

\title{Active doping controls the mode of failure in dense colloidal gels}

\author[a,b,1]{Tingtao Zhou}
\author[a,b,1]{John F. Brady} 

\affil[a]{Division of Chemistry and Chemical Engineering, California Institute of Technology}
\affil[b]{Division of Engineering and Applied Science, California Institute of Technology}

\leadauthor{Zhou} 

\significancestatement{
Adaptive materials that can alter their mechanical properties in response to an external stimulus are highly desirable. Specifically, yielding of disordered materials is a nonlinear process governed by their complex free energy landscape. 
Using Active Brownian Dynamics simulations, we demonstrate precise control of colloidal gel structure, elasticity and ductility by tuning the properties of a remarkably small fraction of embedded active particles, specifically their propulsive swim force. Our findings show that active doping is an effective approach to modulating the mechanical properties of the host material, providing guidance for designing novel reconfigurable materials.
}

\authorcontributions{T.Z. and J.F.B. designed research, analyzed data, performed research, and wrote the manuscript. }
\authordeclaration{The authors declare no conflict of interest.}

\correspondingauthor{\textsuperscript{1}To whom correspondence should be addressed. E-mail: edmondztt@gmail.com; jfb@cheme.caltech.edu}


\begin{abstract}
Mechanical properties of disordered materials are governed by their underlying free energy landscape.  
In contrast to external fields, embedding a small fraction of active particles within a disordered material generates non-equilibrium internal fields, which can help to circumvent kinetic barriers and modulate the free energy landscape. 
In this work, we investigate through computer simulations how the activity of active particles alters the mechanical response of deeply annealed polydisperse colloidal gels. We show that the 'swim force' generated by the embedded active particles is responsible for determining the mode of mechanical failure, i.e., brittle vs. ductile. We find, and theoretically justify, that at a critical swim force the mechanical properties of the gel decrease abruptly, signaling a change in the mode of mechanical failure. 
The weakening of the elastic modulus above the critical swim force results from the change in gel porosity and distribution of attractive forces among gel particles, while below the critical swim force, the ductility enhancement is caused by an increase of gel structural disorder. Above the critical swim force, the gel develops a pronounced heterogeneous structure characterized by multiple pore spaces, and the mechanical response is controlled by dynamical heterogeneities. We contrast these results with those of a simulated monodisperse gel that exhibits a non-monotonic trend of ductility modulation with increasing swim force, revealing a complex interplay between the gel energy landscape and embedded activity.
\end{abstract}

\dates{This manuscript was compiled on \today}
\doi{\url{www.pnas.org/cgi/doi/10.1073/pnas.XXXXXXXXXX}}

\begin{document}
\setboolean{displaywatermark}{false}

\maketitle
\thispagestyle{firststyle}
\ifthenelse{\boolean{shortarticle}}{\ifthenelse{\boolean{singlecolumn}}{\abscontentformatted}{\abscontent}}{}

\dropcap{B}iological matter has the amazing capability of adapting to its environment. For instance, cells alter their stiffness and viscosity in response to different substrate rigidities~\cite{lo2000cell} by polarizing their actin cytoskeleton network~\cite{doss2020cell}.
Adaptations generally render more robust response to environmental changes and enhance the chances of survival for living organisms.
Inspired by biology, can one endow passive synthetic materials with similar controlled mechanical adaptivity? In this work we consider a simple system of passive colloidal particles where a small fraction of them can be activated by external stimuli to self-propel, i.e., become active Brownian particles (ABPs)~\cite{fily2012athermal}.
The persistent motion of ABPs distinguishes them from passive Brownian particles and can be understood as a source of mechanical force or stress, i.e. the swim stress~\cite{takatori2014swim,omar2020microscopic}. In fact, it is possible to extract mechanical work from self-propelling particles and transmit free energy to the surrounding or to other components of the system~\cite{angelani2009self,reichhardt2017ratchet,ro2022model}. This mechanical perspective suggests that embedded activity may be harnessed to modulate the mechanical properties of a host material.

Recent advances in colloidal physics and chemistry provide the ability to engineer colloidal particle shapes~\cite{damasceno2012predictive} and interaction potentials~\cite{wang2012colloids}, which have led to the self-assembly of various structured colloidal materials with different symmetries and hierarchies~\cite{vlasov2001chip,liu2014orientationally,engel2015computational,wang2017colloidal,he2020colloidal}. 
As an alternative to this passive self-assembly, active doping -- namely embedding a small fraction of active particles into the host material -- can provide a non-equilibrium route for directed self-assembly that may overcome  kinetic arrest and frustration in passive self-assembly. 
For example, adding a small amount of active particles can promote the 
crystallization of colloidal hard spheres~\cite{ni2014crystallizing} or attractive particles~\cite{mallory2020universal}, and anneal defects in dense  colloidal monolayers~\cite{ramananarivo2019activity,saud2023microdynamics}. In disordered colloidal gels, Omar et al.~\cite{omar2018swimming} showed that the dynamical free energy landscape can be engineered by changing the self-propulsive swim force and the length and time scales of the persistent motion of active particles, resulting in widely different gel structures.

The mechanics and rheology of dense gels and glasses are governed by the inherent structures of the energy landscape~\cite{sastry2001relationship,sastry1998signatures,heuer2008exploring} and by dynamical heterogeneities~\cite{berthier2011dynamic,berthier2011overview}. Linear response properties in disordered hard-sphere packings are known to relate to the glass transition~\cite{wittmer2013shear,yoshino2014shear} and follow universal scalings~\cite{goodrich2016scaling} at the jamming point.
As a nonlinear response to external loading, yielding and plasticity can be viewed as stress and thermal activation processes on a complex energy landscape~\cite{cao2019potential}.
The ductility of yield-stress colloidal materials, such as cement~\cite{masoero2012nanostructure,ioannidou2016mesoscale}, is a property that determines the integrity of human-made  structures, such as buildings and bridges, and thus is relevant to their life-time and durability. At the same time,  structural features, such as pore network connectivity and topology, have fundamental implications for multiple degradation mechanisms of heterogeneous materials~\cite{zhou2019multiscale,zhou2020freezing,monfared2020effect}.
It remains to be understood how the internal energy sources in active matter interplay with the underlying energy landscape of the host material and consequently modulate its structural and mechanical properties.

Recent experiments on dilute gels~\cite{szakasits2019rheological,saud2021yield} observed a dramatic threefold change in the yield stress due to a tiny fraction (0.1~\%) of active particles disrupting the gel structure. The activity-induced structural changes were long-lasting even after activity was switched off~\cite{wei2023reconfiguration}, demonstrating a strong impact of activity on the aging behavior of dilute gels. A wide range of practical colloidal materials is dense, glassy and much stiffer, such as cement paste. 
How does activity alter the mechanical behaviors of such dense gels?

In this work, we consider a simple system of passive colloidal particles with a small fraction ($\sim10\%$) of embedded ABPs. The ABPs self-propel at a fixed swim force.
We develop an active Brownian dynamics (BD) model to numerically demonstrate precise control of the structure and mechanical properties of dense disordered colloidal gels by altering the activity of the embedded active particles. Specifically, we show a fine control of the gel's brittle-ductile transition by varying the swim force, accompanied by a nonlinear change of the gel pore morphology and elastic properties. 
\change{Our results demonstrate a protocol for modifying colloidal gels in response to external loading so that they deform in a ductile manner instead of catastrophic brittle failure.}
These results pave the way to adaptive materials design by active doping.

\begin{figure*}[tbh]
\centering
\includegraphics[width=.95\linewidth]{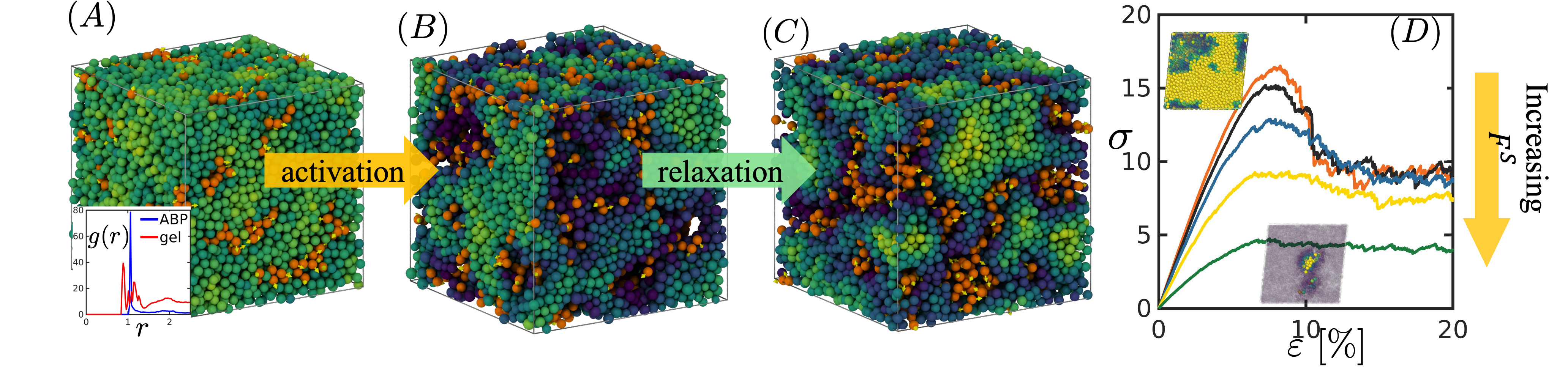}
\caption{
\revise{Active doping protocol and key effects. (A) A typical initially annealed configuration. Green for polydisperse passive gel particles, darker particles have larger coordination numbers. Orange for ABPs. Yellow arrows denote the randomized initial orientation of the ABPs. The fraction of ABPs in each sample is about $10\%$. The inset shows $g(r)$ for the gel-gel particles and ABP-ABP particles, indicating a micro-phase separation of ABPs. (B) Due to activation of ABPs, the gel undergoes volume expansion and structrual alterations. (C) After activity is turned off, the gel relaxes and exhibits irreversible structural changes. (D) Simple shear stress-strain curves for gels treated by the activation protocol (A-C). The mode of failure changes from brittle to ductile yielding as swim force $F^S$ increases. The insets show typical yielding behaviors (shear banding v.s. localization) for polydisperse gels activated at different $F^S$.}}
\label{fig:fig1}
\end{figure*}

\section*{Model of deeply annealed gels doped with active particles}

\revise{The particle size distributions in dense colloidal gels vary; for instance, cement paste consists of a range of particle sizes~\cite{masoero2012nanostructure,ioannidou2016mesoscale}, while colloidal crystals~\cite{hueckel2020ionic} can be made monodisperse or bi-dispersed. We will focus on the polydisperse gels (See methods) and contrast them with the monodisperse ones later.}
We model dense active-doped gels as random  packings of spheres, where only $\phi_A=$10\% of the particles are Active Brownian Particles (ABPs), as shown in Fig.~\ref{fig:fig1}(A). 
ABPs' motion is overdamped and a particle moves at a velocity $\bm{U}$ that satisfies the force balance 
\begin{equation}
-\zeta \bm{U} + \zeta \bm{F}^S + \bm{F}^{B} + \bm{F}^{P} = 0, 
\end{equation}
where $\zeta$ is the fluid drag, and $\bm{F}^S=\zeta U_0\bm{q}$ is the self-propulsive swim force, with ${U}_0$ the swim speed in the direction $\bm{q}$.
$\bm{F}^P$ is the interparticle force and $\bm{F}^B$ is the random Brownian force. An ABP is also subject to a random torque that changes the orientation $\bm{q}$  over the reorientation timescale $\tau_R$ (see Materials and Methods). 
The passive colloidal particles follow the same dynamics except that they do not self-propel ($\bm{F}^S=0$.).

Activity of the ABPs will be switched on and then off, similar to recent experiments for quasi-2D gels~\cite{wei2023reconfiguration}.
The passive gel particles strongly attract each other (attraction potential well depth $\epsilon=10~k_BT$; see Materials and Methods), but the ABPs are not attracted to the gel particles (or each other) and behave as soft spheres. 
Before switching on the activity of the ABPs (their swim force is $\bm{F}^S=0$ at this moment), the active-doped gel samples are annealed to low effective temperature via the swap Monte Carlo algorithm~\cite{grigera2001fast,ninarello2017models}. As a result, we obtain stable dense gels that are brittle in nature~\cite{ozawa2018random}. 
After annealing to a low temperature ($\frac{\epsilon}{k_BT}\gg1$), the interaction energy dominates over mixing entropy and favors micro-phase separation where the ABPs aggregate to form pockets naturally, as shown in Fig.~\ref{fig:fig1}. 

\section*{Protocol for activated gels: nonlinear dynamics of constant pressure activation}

The activation process proceeds by turning on the active swim force of the ABPs to a finite value $F^S$ instantaneously, with randomized initial orientations $\bm{q}$ and reorientation time $\tau_R=2~\tau_B$, where $\tau_B$ is the Brownian timescale (see Methods). 
Activation causes the gel to expand, and, as in an experiment, the gel expands at constant pressure. 
The pressure, \change{or the active stress} of the gel is the sum of the kinetic pressure $nk_BT$ and the virial part $\Pi^{a}=\frac{1}{3V}\sum_{\alpha}\bm{x}_\alpha\cdot\bm{F}_{\alpha}^t$.
Here, $\bm{F}_{\alpha}^t$ is the total force exerted on particle $\alpha$ from the other passive and active particles, including the swim force, for which we use the impulse formula:
$\bm{x}_\alpha\cdot\bm{F}_{\alpha}^S=\tau_R\bm{U}_\alpha\cdot\bm{F}^S_\alpha$.

\begin{figure*}[ht]
\centering
\includegraphics[width=.95\linewidth]{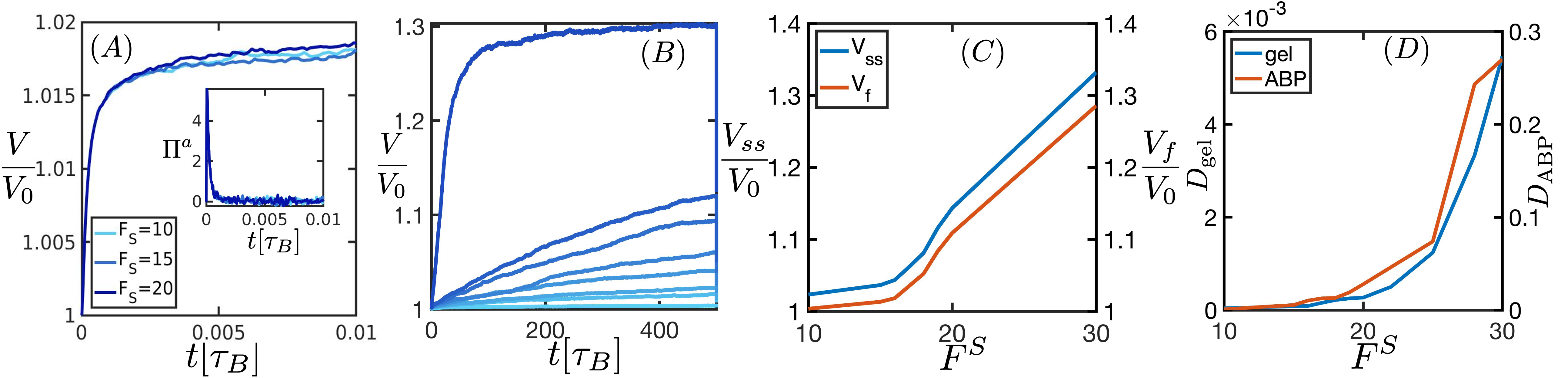}
\caption{Nonlinear dynamics during the activation processes. 
(A) \change{Typical volume expansions at the early stage immediately after the ABPs' activity is turned on. Inset shows the instantaneous active stress is large at $t=0$ but quickly decays.
(B) Quasi-static gel volume expansion during activation; $F^S$ increases from 10 to 30~$k_BT/a$ as line colors become darker. 
(C) Both the steady state and final gel volumes increase as $F^S$ increases.
(D) The self-diffusivity of the gel particles and ABPs during activation.}}
\label{fig:fig-dynamics}
\end{figure*}

Once the ABPs are activated, the unit cell expands isotropically according to the non-equilibrium constant pressure barostat~\cite{wang2015constant}
\begin{equation}
\frac{\dot{L}}{L} = \frac{1}{\kappa_T} \left(
n k_B T + \frac{1}{3}\left<\bm{x}\cdot\bm{F}^t\right> - \Pi_0
\right),
\end{equation}
where $\kappa_T$ is an isothermal bulk viscosity of the gel and $\Pi_0$ is the ambient pressure set to balance the thermal pressure before activating the ABPs.
\change{Immediately after the activity is turned on, the active stress is large and positive (Fig.~\ref{fig:fig-dynamics}A inset), leading to an instantaneous volumetric expansion (Fig.~\ref{fig:fig-dynamics}A) that is far from equilibrium in nature. The excess stress quickly decays due to the volume expansion, and the entire system then expands quasi-statically over timescales longer than the Brownian time}, as shown in Fig.~\ref{fig:fig-dynamics}(B).

To make a rough estimate of the effect of activation,
a homogeneous gel will rupture at a pressure $P_C\sim \frac{N}{V}Z_{\text{eff}}'F_{\text{mean}}a$ \revise{for $N$ gel particles of diameter $a$ occupying volume $V$. }
The typical attractive force between gel particles $F_{\text{mean}}\sim10 k_BT/a$, originating from their interaction potential. Here, $Z_{\text{eff}}'\sim O(1)$ is the effective per particle nearest neighbor pairs across a plane, and $a$ is the average particle size.
The homogeneous collisional pressure from the ABPs is $\Pi_c\sim \phi_A F^S a \frac{N}{V}$, which gives the critical value to rupture the gel through isotropic expansion: $F^S\sim Z_{\text{eff}}' F_{\text{mean}}/\phi_A\sim10^2 k_BT/a$. Hence, within a reasonably mild activity range, the gel will not be torn apart by the ABPs.

On the other hand, at a local scale the ABPs can break apart passive particle pairs and penetrate into the gel. The critical swim force $F^S_c$ to break  a pair of attractive passive particles  can be estimated by considering the energy of quasi-statically breaking a bond:
\begin{equation}
F^S_c \sim \frac{\epsilon}{a} Z_{\text{eff}},
\end{equation}
where $Z_{\text{eff}}$ is the effective coordination number of passive gel particles. Simulation results give $F^S_c\sim15 k_BT/a$, or $Z_{\text{eff}}\approx1.5$, which is reasonable. This critical swim force separates qualitatively different regimes of the gel structure and mechanics, as discussed below.

After steady state, we turn off the activity of ABPs ($F^S=0$), and let the system relax to the initial ambient pressure $\Pi_0$. The total volume shrinks as the active collisional pressure is eliminated. The final volume is always larger than the initial value and grows as swim force increases, as shown in Fig.~\ref{fig:fig-dynamics}(C), indicating irreversible structural changes.  We then quantify the structural changes compared to the initial annealed state.

\section*{Activity-modulated elasticity}

To determine the effect of active doping on the mechanical properties of the gel, we perform athermal quasi-static (AQS) simple shear simulations on the final gel configurations. During shear, the ABPs remain deactivated ($F^S=0$). 
Below, we denote gel samples by the $F^S$ (in units of $k_BT/a$) values of the ABPs during the activation processes.

\begin{figure*}[thb]
\centering
\includegraphics[width=.95\linewidth]{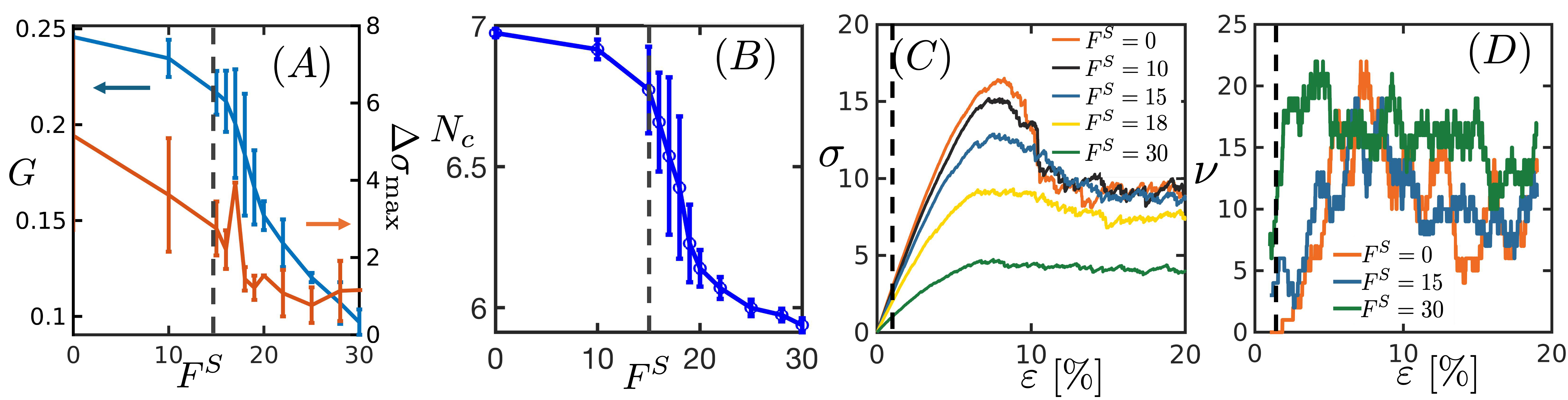}
\caption{Mechanical properties of the gels after activation-deactivation processes at different swim forces.
(A) Both elastic modulus ($G$, blue line, normalized by $G_{\text{FCC}}$) and maximum stress drop size max($\sigma$) decrease with swim force.
(B) Mean contact number per particle $N_c$ as a function of $F^S$. \revise{Vertical dash lines in (AB) indicate the predicted critical swim force $F^S_c\sim15k_BT/a$.}
(C) Stress-strain curves for the initial gel and activated gels during athermal-quasi-static shear tests, each curve is averaged over 5 samples. (D) Typical stress drop frequencies (moving averaged) during shear. \revise{Vertical dash lines in (CD) indicate $\varepsilon=1\%$.}
}
\label{fig:fig4-mechanics}
\end{figure*}

First, we examine the linear response of gels activated at different $F^S$. We quantify the shear modulus using the initial slope of the stress-strain curve at shear strain $\varepsilon<1\%$. 
The initial gels are strong compared to the previous experimental dilute gels~\cite{szakasits2019rheological,saud2021yield} reflected by their relatively high shear modulus $G$. As shown by the blue line in Fig.~\ref{fig:fig4-mechanics}(A), $G(F^S=0)\approx0.25~G_{\text{FCC}}$, where $G_{\text{FCC}}$ is the modulus of a FCC crystal calculated from the same AQS protocol.
For these inactivated gels, their elastic regime is quite ideal, evidenced by the moving-averaged frequency of stress drops as shear strain increases, as shown by the orange line in Fig.~\ref{fig:fig4-mechanics}(D). Very few stress drop appear at the early stage of small $\varepsilon<1\%$.

After activation, the shear modulus $G$ is weakened as $F^S$ increases. 
Noticeably, the decrease in $G$ is nonlinear in $F^S$, with a drastic decline around $F^S\sim F^S_c$.
Although $G$ decreases rapidly around $F^S_c\sim 15~k_BT/a$, it remains finite and the same order of magnitude even at $F^S=30~k_BT/a$ (from initial $G_0\sim 0.25~G_{\text{FCC}}$ to $G_0\sim 0.1~G_{\text{FCC}}$). Hence, $F^S_c$ is below the force necessary to ``fluidize'' the gel, where the shear modulus decreases to zero. 
According to the estimate in the last section, ABPs with swim forces above $F^S_c$ can break local bonds, and cages formed during gel annealing can be broken in this regime.
As a result, the elastic regime of highly activated gels deviates from the ideal scenario, and becomes populated with microscopic plastic deformations even at small strain $\varepsilon<1\%$. These localized irreversible deformations are recorded in the higher initial frequency of stress drops, as shown by the blue and green lines in Fig.~\ref{fig:fig4-mechanics}(D); the system loses rigidity as $F^S$ increases.

The structural origin of the rigidity loss can be understood from the perspective of the Cauchy-Born theory of amorphous solids~\cite{alexander1998amorphous,zaccone2009elasticity}. To clarify this, we define the mean number of contacts per particle $N_c$, as shown in Fig.~\ref{fig:fig4-mechanics}(B) (see Materials and Methods). 
For gels below $F^S_c$, $N_c$ stays large, which is consistent with the rigid gel response to loading. For gels activated at $F^S>F^S_c$, $N_c$ decreases rapidly, indicating the onset of loss of isostaticity~\cite{maxwell1864calculation}.
As one expects from the Cauchy-Born theory, the shape of $G(F^S)$ resembles that of $N_c(F^S)$, comparing Fig.~\ref{fig:fig4-mechanics}(A) and (B). 
The correlation between $N_c$ and $G$ observed here is consistent with previous experiments on passive colloidal gels~\cite{whitaker2019colloidal,hsiao2012role}. Overall, our findings here show that tuning $F^S$ can control the gel network topology and elastic modulus.

\section*{Activity-induced brittle-to-ductile transition}

Preluded by the increasing abundance of microscopic plastic deformations at small strain, 
for large deformations that yield the gels, the stress-strain responses undergo a continuous change from brittle to ductile behavior as $F^S$ increases, as shown by Fig.~\ref{fig:fig4-mechanics}(A) red line and (C). 
The yield point is taken as when the macroscopic shear stress saturates or starts to decrease as shear strain increases. The yield strain is always around $\varepsilon\sim 8\%$, regardless of different $F^S$ values. The yield stress, however, decreases continuously as $F^S$ increases.
To quantify the gel ductility, we measure the \change{stress drops} $\Delta\sigma$ (see Materials and Methods) during the simple shear processes. As expected, the maximum stress drop $\Delta\sigma_{\text{max}}$ coincides with the yielding transition and the magnitude of $\Delta\sigma_{\text{max}}$ is used as a scalar measure of brittleness.

The inactivated gels are quite brittle, with a significant stress overshoot in their stress-strain curves, as shown by the orange curve of Fig.~\ref{fig:fig4-mechanics}(C). Correspondingly, $\Delta\sigma_{\text{max}}$ is largest at $F^S=0$.
For gels activated at different $F^S$, $\Delta\sigma$ follow similar truncated power-law distributions (SI Fig.~1), but $\Delta\sigma_{\text{max}}$ decreases as $F^S$ increases, as seen in Fig.~\ref{fig:fig4-mechanics}(A).
In contrast to the drastic decrease of $G$ around $F^S_c$, $\Delta\sigma_{\text{max}}$ decreases mildly with increasing $F^S$.
Meanwhile, the total number of stress drops, or plastic events, increases when $F^S>F^S_c$, as shown by the total areas under different curves in Fig.~\ref{fig:fig4-mechanics}(D). 
Overall, these statistics of plastic events indicate that the gels become more ductile as $F^S$ increases. 

\begin{figure}[htb]
\centering
\includegraphics[width=.95\linewidth]{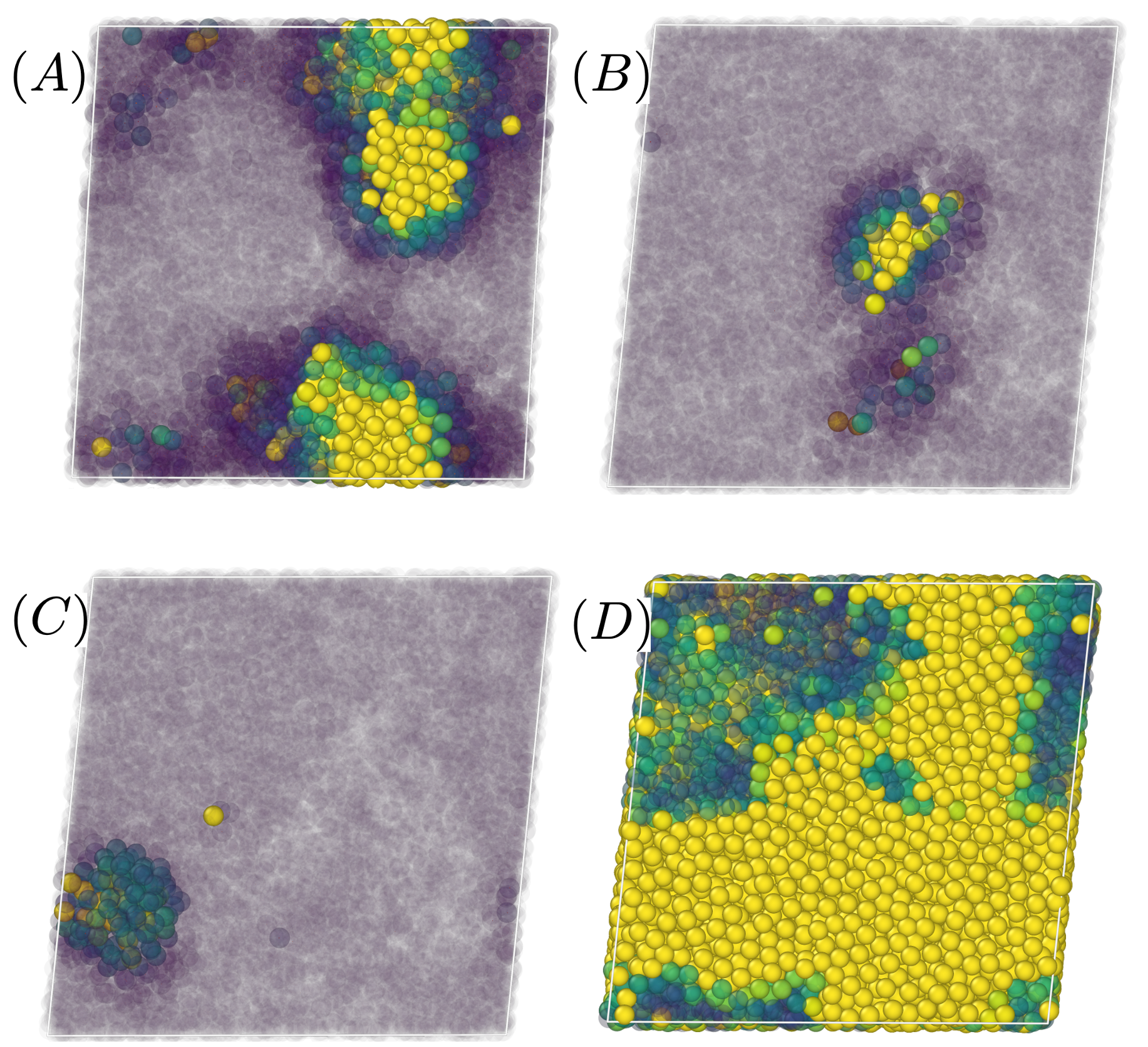}
\caption{Typical shear transformations for $F^S=20$ \change{during stress drop events at shear strains of} (A) $\varepsilon=5.47\%$, (B) $\varepsilon=9.90\%$, (C) $\varepsilon=10.16\%$ and (D) shear banding for $F^S=0$ at $\varepsilon=10.15\%$. No shear banding is observed for $F^S=20$. \change{Yellow particles have large non-affine displacements $D_{\text{min}}^2$ (>1 particle diameter)} during the stress drop event. \revise{Particles with small $D_{\text{min}}^2$<1 are rendered transparently to highlight the shear transformation zones.}
}
\label{fig:fig5-nad}
\end{figure}

We notice that the $\Delta\sigma_{\text{max}}$ at $F^S=30~k_BT/a$ is still finite ($\sim O(1)$), raising the question of whether the yielding behavior can be classified as truly ductile or rather a series of small yet brittle avalanches. Current continuum models give different predictions on the nature (discontinuous or continuous) of the brittle-ductile transition~\cite{ozawa2018random,singh2020brittle,pollard2022yielding,rossi2022finite}.
To further clarify our quantification of gel ductility, we examine the spatial distributions of shear transformations during shear. For all gels activated at $F^S>F^S_c$, there is no shear banding that percolates the entire sample; in contrast, such percolating shear bands always show up in the deeply annealed gels without activation (Fig.~\ref{fig:fig5-nad}A-D).
The percolation behavior of shear transformation zones is consistent with the observed trends of $\Delta\sigma_{\text{max}}$ and the avalanche frequency $\nu$ (see Materials and Methods) with respect to $F^S$. 
The absence of shear bands suggests that the gels activated above $F^S_c$ undergo truly ductile yielding. 
Below $F^S_c$, the characteristic behaviors of $\Delta\sigma_{\text{max}}$ and $\nu$ indicate that the gels continuously enhance ductility as $F^S$ increases.

\begin{figure*}[bh]
\centering
\includegraphics[width=.95\linewidth]{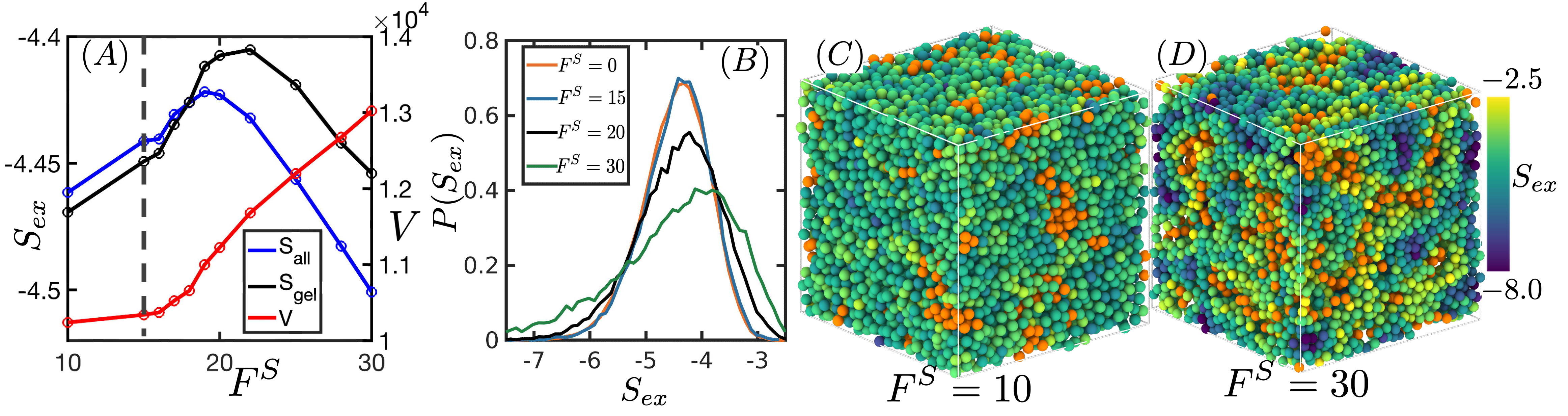}
\caption{Local degree of structural disorder measured by the excess entropy $S_{\text{ex}}$. (A) the excess entropy (left y-axis) averaged over all particles ($S_{all}$ blue line) and gel particles ($S_{gel}$ black line), and the final total volume $V$. The initial total volume is $V_0=1.03\times10^4~a^3$. \revise{Vertical dash line indicates the predicted critical swim force $F^S_c\sim15k_BT/a$.} (B) Probability distribution of $S_{\text{ex}}$. (C)(D) show snapshots of the final gels activated at $F^S=10,30$, respectively, with gel particles colored by their local excess entropy and the ABPs colored in orange. }
\label{fig:fig-sex}
\end{figure*}

\section*{Activity-modulated structural entropy change and its connection to brittle-to-ductile transition}
To understand the underlying structural changes responsible for the enhanced ductility, we examine the degree of disorder of the passive gel particle configurations by calculating the local excess entropy $S_{\text{ex}}=S-S_{\text{ig}}$~\cite{rosenfeld1977relation,dyre2018perspective}  (see Materials and Methods) around each particle and average over the sample, where $S_{\text{ig}}$ is the ideal gas entropy. 
Below $F^S_c$, the gel structure becomes more disorder as $F^S$ increases, as seen by the increase of $S_{\text{ex}}$ with $F^S$ shown in Fig.~\ref{fig:fig-sex}(A).
The enhancement in ductility accompanied by the increase of disorder is consistent with recent observations in other disordered materials, such as silica or metallic glasses~\cite{tang2021energy}. In that study, structural disorder was assumed to originate from irradiation. In fact, ABPs share similarities with irradiation bombardment in terms of their ballistic motions at short times.
$S_{\text{ex}}$ saturates when $F^S$ is slightly above $F^S_c$ and then decreases with increasing $F^S$, indicating that the local arrangement of gel particles is more ordered, even though the gel is more ductile.

\begin{figure*}[htb]
\centering
\includegraphics[width=.95\linewidth]{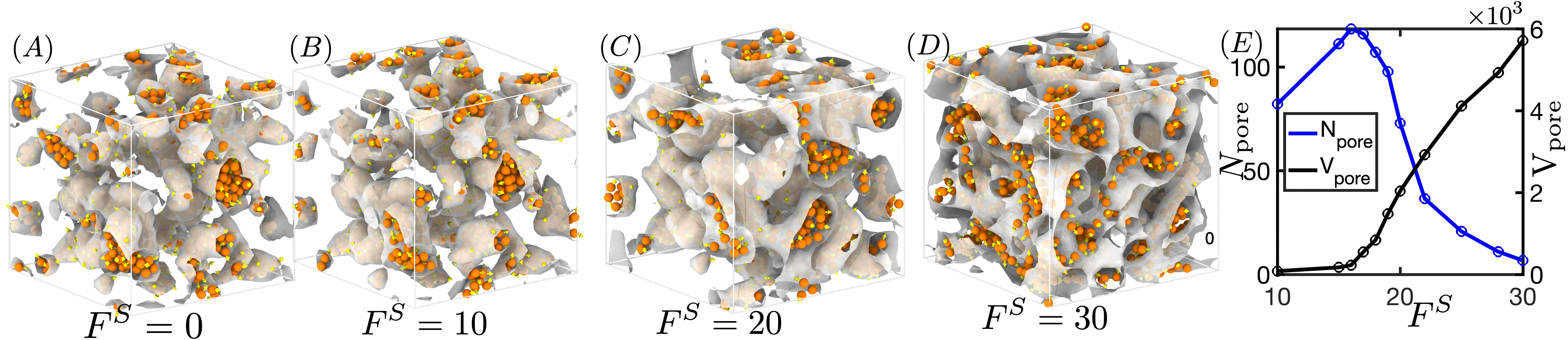}
\caption{Structural changes of the passive gel induced by the ABPs. (A-D) compare the final \revise{pore} morphology of the passive gels \revise{($F^S=0$) with those} activated at $F^S=10$, $F^S=20$, and $F^S=30$. ABPs are orange spheres with yellow arrows indicating their orientation, gel particles not shown. (E) shows the total number of pores in the sample $N_{\text{pore}}$ (blue line, left y-axis) and the total volume of pore space $V_{\text{pore}}$ (black line, right y-axis), as a function of the swim force of ABPs.
}
\label{fig:fig-pore}
\end{figure*}

\begin{figure}[htb]
\centering
\includegraphics[width=.95\linewidth]{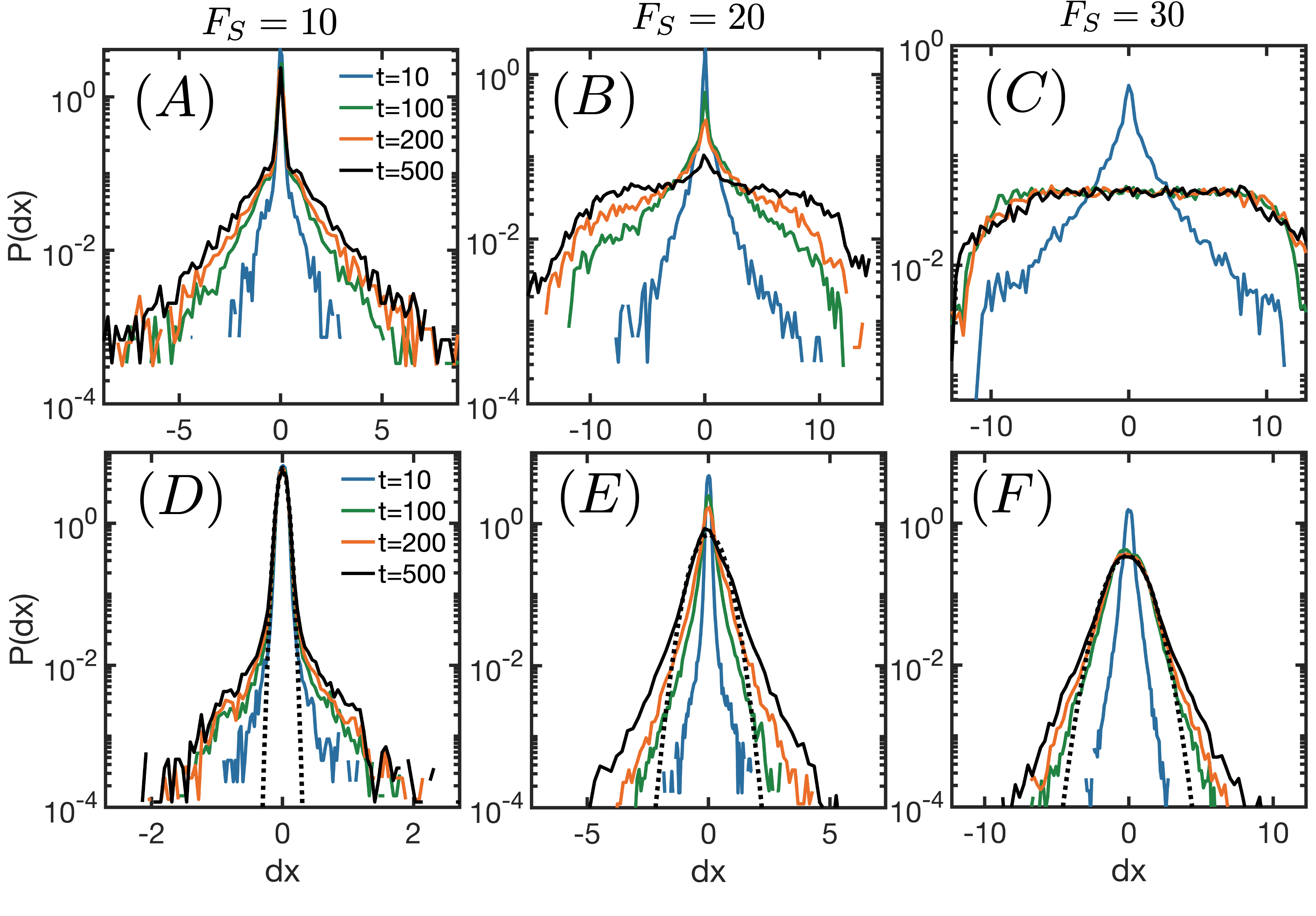}
\caption{
\revise{Distribution of particle displacements $P(dx)$ at (A,D) $F^S=10$, (B,E) $F^S=20$ and (C,F) $F^S=30$ for the polydisperse gels. (A-C) for ABPs, (D-F) for passive particles. Black dash lines in (D-F) show the Gaussian part of $P(dr)$ at $t=500\tau_B$. As $F^S$ increases, the long-time behavior of $P(dr)$ correlation deviates more from Gaussian. In contrast, $P(dr)$ for the passive particles becomes more Gaussian for larger $F^S$: the kurtosis of $P(dr,t=500\tau_B)$ are $k={\left<dr^4\right>}/{\left<dr^2\right>^2}= $ 29.6, 19.7, 8.7 for $F^S=10,20,30$ respectively.}
}
\label{fig:fig-pdr}
\end{figure}

\begin{figure}[htb]
\centering
\includegraphics[width=.95\linewidth]{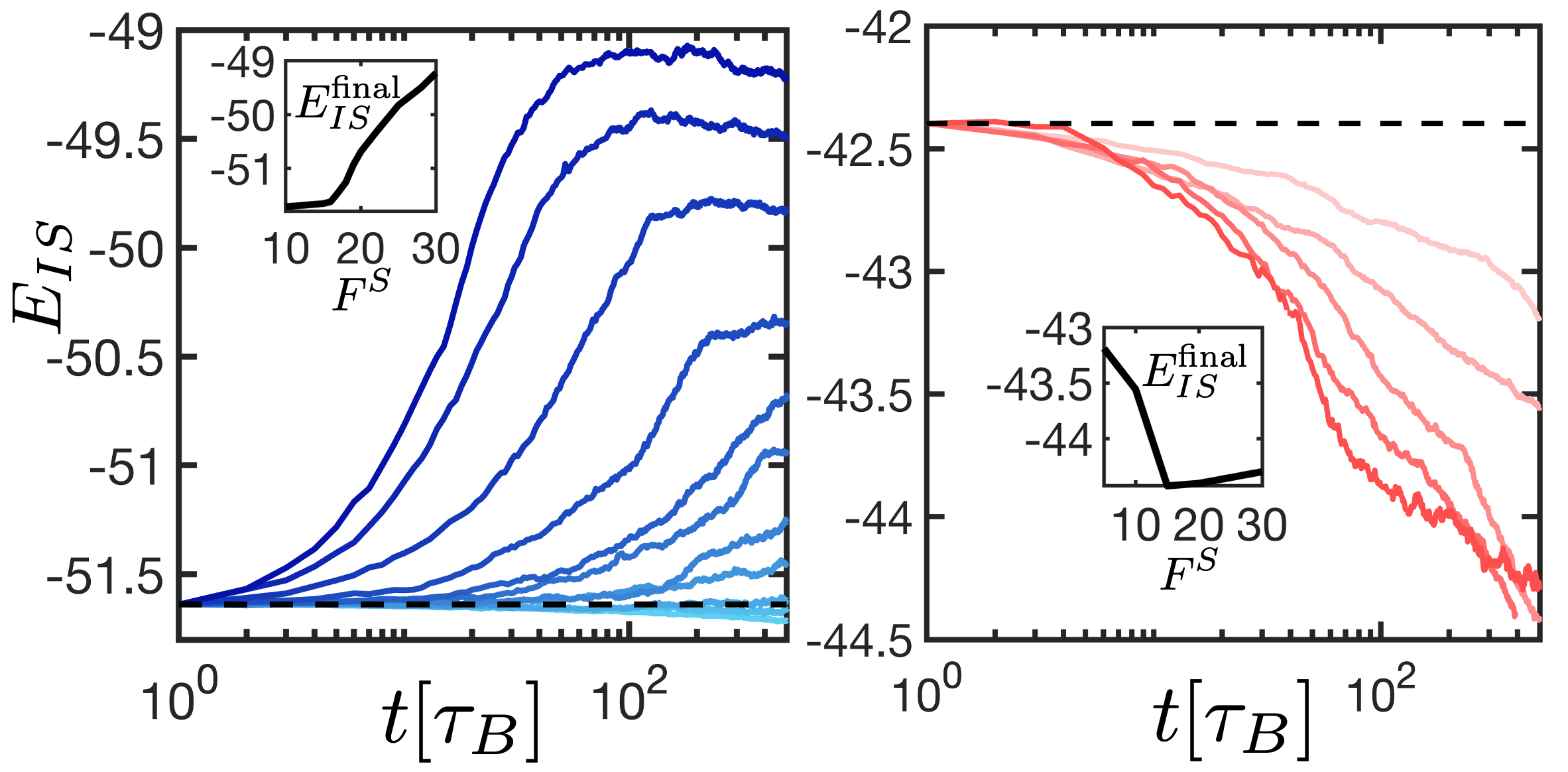}
\caption{\revise{Inherent structure energy evolution as activation proceeds. Colors from light to dark blues (reds) correspond to increasing $F^S$. (A) Polydisperse gels. $E_{\text{IS}}$ increases with $F^S$ for $F^S>F_c$. (B) Monodisperse gels. $E_{\text{IS}}$ decreases with increasing $F^S$ and saturates above $F_c$. 
}}
\label{fig:fig-EIS}
\end{figure}

\section*{Activity-modified pore networks enhance dynamical heterogeneity}
To rationalize the reversed trend of $S_{\text{ex}}$ at large $F^S$, we notice that above $F^S_c$, the gel pore volume increases significantly with $F^S$ (Fig.~\ref{fig:fig-pore}D). 
The irreversible volume expansion induced by activity increases the internal pore space volume $V_{\text{pore}}$, as shown in Fig.~\ref{fig:fig-pore}(D). 
Below the critical swim force, both $V_{\text{pore}}$ and number of pores $N_{\text{pore}}$ increase mildly with $F^S$. In this regime, the ABPs are not able to penetrate into the gel and disrupt its structure, but rather they locally strain the gel and create a multitude of small pores. Here, $S_{\text{ex}}$ remains a good indicator for local degree of order in the particle configuration and gradually increases.

However, above $F^S_c$ the ABPs penetrate through the gel backbone, enhancing the pore connectivity. As a result, $N_{\text{pore}}$ decreases quickly, while $V_{\text{pore}}$ increases rapidly, as shown in Fig.~\ref{fig:fig-pore}(C).  
Typical final gel pore structures are shown in Fig.~\ref{fig:fig-pore} (A-C) at swim forces below, near and above the critical force $F^S_c$.
Hence, the gel develops a larger pore network and exhibits stronger structural heterogeneities, as supported by the broadened distribution of $S_{\text{ex}}$ in Fig.~\ref{fig:fig-sex}(B). 
In this regime, the decrease of $S_{\text{ex}}$ emphasizes only small nucleations of crystalline ordered clusters, but does not reflect the larger scale pore topology.
A closer examination of the most orderly arranged particles with $S_{\text{ex}}<{\bar{S}_{\text{ex}}} - \sigma_{S}$, 
where $\bar{S}_{\text{ex}}$ is the mean of $S_{\text{ex}}$ and $\sigma_{S}$ is the standard deviation, shows that these particles do not form a percolating backbone to sustain the globally applied external load; hence, they do not dominate the gel's response to shear. Instead, the enhanced overall porosity and heterogeneity now take over and increase gel ductility.
Correspondingly, the system behaves dynamically more heterogeneous, as shown by the multitude of localized shear transformations with small stress drops before yielding (Fig.~\ref{fig:fig5-nad}).

\begin{figure*}[htb]
\centering
\includegraphics[width=.95\linewidth]{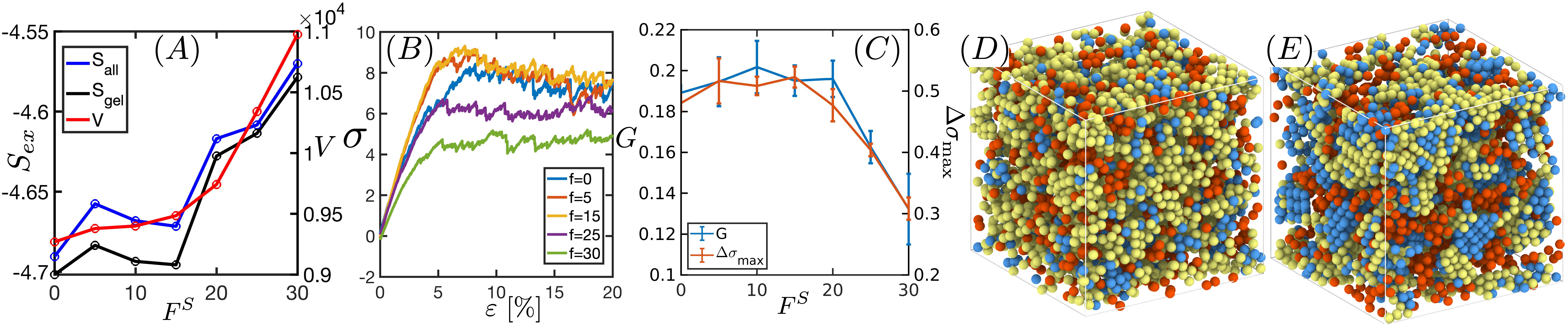}
\caption{Active-doped monodisperse gels. (A) Local excess entropy $S_{\text{ex}}$ averaged over all particles (blue line, left y-axis) and passive particles only (black line, left y-axis), and the final system volume $V$ (red line, right y-axis). (B) Averaged stress-strain curves during athermal quasi-static shear. (C) Elastic moduli $G$ (blue line, left y-axis, normalized by $G_{\text{FCC}}$) and the maximum stress drop size $\Delta\sigma_{\text{max}}$ (red line, right y-axis). Typical snapshot for final gel configurations at (D) $F^S=10$ and (E) $F^S=30$, where ABPs are colored orange, passive particles with local FCC order colored blue, HPC order colored yellow. Other passive particles, i.e. disordered ones, are not displayed. 
}
\label{fig:fig-mono}
\end{figure*}

\section*{Activity drives the energy landscape}
From the perspective of inherent structures in glassy states, the changes in mechanical properties discussed above should be reflected in the different local environments on the energy landscape.
Indeed, during activation, the gel particles significantly enhance their mobility as $F^S$ increases, as shown by the self-diffusivities in Fig.~\ref{fig:fig-dynamics}(D). Their steady-state mean-squared-displacements (MSD) exceed a particle diameter when $F^S>F^S_c$ (SI Fig.1), indicating cage-breaking events during activation, consistent with the bond-breaking argument and the observed change in inter-particle contacts. 
\revise{Non-Gaussianity of displacement distribution for active particles in polymer networks has been extensively studied~\cite{kim2022active,yadav2023dynamics,kumar2023dynamics,ben2015modeling}.}
\revise{
To better illustrate the dynamics of the particles during activation, we compute the self part of the van Hove correlation function $P(dr)$, as shown in Fig.~\ref{fig:fig-pdr}.
$P(dr)$ at long times for ABPs deivates from Gaussian more as $F^S$ increases, as one may expect. Interestingly, the long time behavior of $P(dr)$ for the passive particles approach closer to Gaussian for larger $F^S$, with decreasing kurtosis of $P(dr,t=500\tau_B)$.
Our observations on the activation dynamics are consistent with the enhanced dynamical heterogeneity in models of active glasses in a constant volume ensemble~\cite{paul2023dynamical}.
}
\revise{To demonstrate the system's trajectories on the energy landscape, we turn off activity at each time point and relax the gels to local potential energy minima at zero temperature and pressure. The inherent structure energy $E_{\text{IS}}$ obtained is shown in Fig.~\ref{fig:fig-EIS}(A). Below $F_c$, $E_{\text{IS}}$ slightly decreases as activation proceeds. For $F^S\gtrsim F_c$, $E_{\text{IS}}$ increases with larger $F^S$.}
Hence, we argue that the embedded activity drives the system to effectively tunnel through energy barriers and explore the landscape. 

\section*{Effects of active doping in monodisperse gels}

To contrast the above results obtained from strong glass formers made of polydisperse particles, we perform the same numerical protocols to 3D gels made of monodisperse particles, again with 10\% ABPs. After thermalization at $1~k_BT$, they form disordered gels similar to the polydisperse case. 

A similar critical value of swim force $F^S_c\approx15$ is observed, where the system volume increases mildly below $F^S_c$ and drastically above $F^S_c$ (Fig.~\ref{fig:fig-mono}A).
However, the effects of activity on the structure and mechanics show qualitative differences from the polydisperse gels. 
The averaged local excess entropy $S_{\text{ex}}$ stays at low values below $F^S_c$ and increases quickly for $F^S>F^S_c$. The increase of $S_{\text{ex}}$ merely reflects the decrease in packing density --- locally particles form clearly visible crystallites, as shown in Fig.~\ref{fig:fig-mono}(D,E), \revise{in contrast to the polydisperse gel in Fig.~\ref{fig:fig-sex}(D)}, and the fraction of crystalline particles increases with $F^S$. Our result is consistent with previous studies showing the activity-induced enhancement of crystalline order in monodisperse packings~\cite{mallory2019activity,ramananarivo2019activity}, distinguishing the role of embedded intrinsic activity from thermal activation. 

Intriguingly, the change in mechanics of the monodisperse gels shows a non-monotonic trend: Below $F^S_c$, the activated gels become tougher, with higher yield stress, as shown in Fig.~\ref{fig:fig-mono}(B), and enhanced elastic modulus $G$ (Fig.~\ref{fig:fig-mono}(C) blue line, normalized by $G_{\text{FCC}}$). The ductility of the gel is marginally reduced (Fig.~\ref{fig:fig-mono}(C) red line). As $F^S$ increases further above $F^S_c$, the gel enters a weakening regime, where both the elastic modulus and yield stress decrease. Meanwhile, it becomes more ductile with smaller yield stress overshoot and $\Delta\sigma_{\text{max}}$. Similar to the polydisperse gels, in this regime, the monodisperse gel's mechanical response to external loading develops multiscale features since the gel expands and significantly changes its pore structure. Local clusters of crystallites are no longer good predictors for the global stress relaxation mechanism, while the globally averaged degree of disorder $S_{\text{ex}}$ is consistent with the enhanced ductility.

We attribute the different trends in mechanical behaviors between activated monodisperse and polydisperse gels to their underlying energy landscapes and different glass-formation propensity.  
The polydisperse gels are prepared at a very stable, deeply annealed state. 
Under the effect of weak embedded activity, statistically the system always finds pathways to less stable glassy states near the current local minimum, \revise{consistent with the trends of $E_{\text{IS}}$
in Fig.~\ref{fig:fig-EIS}(A). }
In the monodisperse gels, nucleation of small crystalline clusters serves as low energy barrier pathways towards the global minimum of a crystal. Hence, even when embedded ABPs perturb the system weakly, it is both thermodynamically and kinetically favorable to first enhance local crystalline order, before a stronger swim force disrupts the gel structure and expands the total volume significantly. 
\revise{The activation process always decreases $E_{\text{IS}}$, driving the system into locally more crystalline states (Fig.~\ref{fig:fig-EIS}B).}
As a result, below $F^S_c$, embedded ABPs reduce the polydisperse gel toughness by increasing disorder, but enhance the monodisperse gel toughness by slightly increasing local crystalline order. Above $F^S_c$, the gel structure is disrupted with many bonds broken, and the effect of increasing pore volume dominates to continuously weaken both polydisperse and monodisperse gels, regardless of the formation of locally ordered clusters.

\section*{Conclusion and Discussions}

In this work we demonstrate the possibility to engineer the mechanical response and mode of failure of colloidal gels by doping them with active swimmers of different activity strength.
With a prototypical computer model of polydisperese colloidal gels doped with Active Brownian Particles, we show that activity dramatically alters the structure and mechanical response of doped colloidal materials. By controlling the particle activity protocol, i.e., on/off timescales, along with the activity strength, $F^S$, we can enhance the degree of irreversible change and \revise{control} the dynamical heterogeneities \revise{(shear transformation zones and shear bands)} present in the mechanical response and dynamics of the passive colloidal gels.
Through theoretical arguments, we predict that there is a critical swim force that an active particle needs to break the bonds between the constituents of the gel. Above this critical swim force, active swimmers enhance the mesoscopic degree of disorder by developing larger pore networks, and thus lead the structure to undergo a brittle-to-ductile transition. This scenario is revealed by our simulations, which show the existence of such a critical force and that the activity-enhanced porous network softens the colloidal structure. Such behavior is reflected by the abrupt decrease of the observed elastic shear modulus and change in degree of disorder, quantified by excess local entropy and pore connectivity measures.

Recent numerical investigations~\cite{tang2021energy} revealed connections between the structural disorder of amorphous materials and their energy landscape via the statistics of local energy barriers.
In this sense, the embedded activity allows passive particles to ``tunnel'' through the energy barriers.
Above the critical swim force $F^S_c$, gels become more porous, with enhanced connectivity.
Porosity and pore connectivity strongly influence the transport properties of porous media~\cite{zhou2019multiscale,pinson2018inferring}, and our results suggest that active doping can be an effective technique to shape the pore networks in disordered materials.

This work is motivated, in part, by the use of dense colloidal materials like cement~\cite{masoero2012nanostructure,ioannidou2016mesoscale,zhou2019multiscale}, where recent experiments have explored the possibility of embedding microbes that can endure the salinity and basic environment in cement paste~\cite{simkiss2012biomineralization,nodehi2022systematic}. 
Typical manufacturing methods for cementitious materials (``curing'') involve complex hydration and aging (``hardening'') processes~\cite{bullard2011mechanisms} over the timescale of days.
We speculate that our results can be applicable to such strong materials, yet the timing and duration to stimulate the embedded active constituents needs further investigation.
\revise{We propose to experimentally investigate the activity-modulated ductility of strong dense gels, potentially made from the aggregation process of attractive colloids, such as by DNA-coatings~\cite{wang2015synthetic}.}

The effects are different for polydisperse and monodisperse disordered gels, reflecting their different underlying energy landscapes.
Our investigation here also provides guidelines for designing reconfigurable materials. For example, we can identify optimal $F^S$, near the critical force $F^S_c$ to maximally enhance the gel ductility in real-time with a minimal weakening of the elastic modulus. In experiments, such tunability can be realized via changing the light intensity, or concentration of fuel chemicals for the active particles.
A possible variation of the protocol is to maintain activity during the shearing process, which may be a useful tool to control material response {\it in situ}.
We hope these results motivate further work, both in continuum theory on the ductility of heterogeneous materials and in experiments of active doped gels in the dense regime.

\matmethods{

\subsection*{Simulation details}

The dynamics of the ABPs and gel particles are integrated according to the over-damped Langevin equation
\begin{equation}
\begin{split}
-\zeta \bm{U} + \zeta \bm{q} U_0 + \bm{F}^{B} + \bm{F}^{P} & = 0, \\
\left<\bm{F}^{B}\right> & = 0,\\
\left<\bm{F}^{B}(0) \bm{F}^{B}(t)\right> &= 2k_BT \zeta \delta(t) \bm{I},\\
\frac{d\bm{q}}{dt} &= \bm{\Omega}^B\times \bm{q},\\
\left<\bm{\Omega}^{B}\right> & = 0,\\
\left<\bm{\Omega}^{B}(0) \bm{\Omega}^{B}(t)\right> &= 2 \delta(t) \bm{I} / \tau_R,\\
\end{split}
\end{equation}
where $\zeta$ is the translational drag coefficient, and
$\bm{q}$ is a unit vector denoting the orientation of the ABP.
$U_0=F^S/\zeta$ is the free space swim speed, which is 0 for passive gel particles.
$\bm{U}$ is the particle velocity. $\bm{F}^B$ is the Brownian force with zero mean and distribution satisfying the fluctuation-dissipation theorem. $\bm{F}^P$ is the force arising from particle-particle interactions.
$\bm{\Omega}^B$ is the random angular velocity resulting from random reorientations and we set $\tau_R=2~\tau_B$ as the persistent runtime. The simulations are in reduced units:  energy is normalized by $k_BT$, length by average particle diameter $a$, and timescale by the Brownian time $\tau_B=\zeta a^2/k_BT$.

The interaction potential between particle $\alpha$ and $\beta$ is
\begin{equation}
V_{\alpha\beta}(r)=C\epsilon\left[
\left(\frac{a_{\alpha\beta}}{r}\right)^n
-\left(\frac{a_{\alpha\beta}}{r}\right)^m
\right],
\end{equation}
where the potential depth is $\epsilon=10~k_BT$, and $n=20$, $m=10$, $C=\frac{n}{n-m}(\frac{n}{m})^{m/(n-m)}=4$. The effective size is the average of the two interacting particle diameters $a_{\alpha\beta}=(a_\alpha+a_\beta)/2$. For passive-passive pairs, the interaction is cut off at $r_p=2.5 \sigma_{\alpha\beta}$. While for either $\alpha$ or $\beta$ being an ABP, the interaction is cut off at $r_a=2^{1/10} a_{\alpha\beta}$ and shifted to only retain the repulsive part.

\revise{For the polydisperse gels,} particle diameters are drawn from a uniform distribution $U(0.8, 1.2)$ \revise{with 20\% fluctuations around the mean}.
\revise{For the monodisperse gels, all particles have diameter $a_P=1.0$. In both cases, }
all ABPs have diameter $a_A=1.0$. Each simulation sample has on average 11000 particles.

\subsection*{Structural characterizations}
The pore surface is determined by the $\alpha$-shape algorithm~\cite{stukowski2014computational}, with a probe sphere radius 0.9, close to the nearest neighbor distance between gel particles.
The local excess entropy for each particle $\alpha$ is estimated by the two-body contribution~\cite{dyre2018perspective}
\begin{equation}
S^\alpha_{ex} \approx S^\alpha_2 = -2\pi\rho_\alpha \int_0^{R} \left[ g_\alpha(r) \ln g_\alpha(r) - g_\alpha(r) + 1
\right] r^2 dr,
\end{equation}
where $\rho_\alpha$ is the local particle number density averaged over a sphere of radius $R=5$ around particle $\alpha$.
The integration is cut off at $R=5$, where the $g_\alpha(r)$ converges to 1.
Including ABPs for computing $S_{\text{ex}}$ shows similar trends, and we focus on the gel $S_{\text{ex}}$ because 
in all the gels (activated or not), the passive particles' rearrangements dominate the stress drops as they constitute the load-bearing backbone of the gels. 
The local crystalline order for the monodisperse gels is identified by the Ackland-Jones bond angle method~\cite{ackland2006applications}.

The mean number of contacts per particle $N_c$ is defined by including all repulsive pairs of particles, as well as attractive pairs whose separation is $r_{12}<a_{12}+\delta r$. 
We show the result for $\delta r=0.01$ in Fig.~\ref{fig:fig4-mechanics}B.
The numeric value of $N_c$ depends on $\delta r$, but the trend with increasing $F^S$ is similar, insensitive to the choice of $\delta r$.

\subsection*{Characterization of stress avalanches}
A stress drop event is defined as a continuous interval of shear strain where the shear stress monotonically decreases. The difference of stress values at the beginning and the end of this stress drop event, $\Delta \sigma$, is denoted as the stress drop magnitude of this event. We then define the frequency of such stress drop events as the number of events per unit amount of incremental strain. For visual clarity, the $\nu(\varepsilon)$ curves in Fig.~\ref{fig:fig4-mechanics}(D) are denoised by moving average with window size $d\varepsilon=0.5\%$.

}

\showmatmethods{} 

\acknow{We acknowledge financial support from the Department of Energy under Grant DE-SC0022966. The simulations in this work are performed on the CFN computing facility at the Brookhaven National Laboratory (proposal ID: GUP-310204). T.Z. thanks S. Chen, H. Row, S. Mallory, D. Frag, X. Wan, and X. Cheng for their helpful discussions.
All study data are included in the article
}

\showacknow{} 

\bibliography{refs}

\begin{thebibliography}{10}

\bibitem{lo2000cell}
CM Lo, HB Wang, M Dembo, Yl Wang, Cell movement is guided by the rigidity of
  the substrate.
\newblock {\em\protect\JournalTitle{Biophysical journal}} \textbf{79}, 144--152
  (2000).

\bibitem{doss2020cell}
BL Doss, et~al., Cell response to substrate rigidity is regulated by active and
  passive cytoskeletal stress.
\newblock {\em\protect\JournalTitle{Proceedings of the National Academy of
  Sciences}} \textbf{117}, 12817--12825 (2020).

\bibitem{fily2012athermal}
Y Fily, MC Marchetti, Athermal phase separation of self-propelled particles
  with no alignment.
\newblock {\em\protect\JournalTitle{Physical review letters}} \textbf{108},
  235702 (2012).

\bibitem{takatori2014swim}
SC Takatori, W Yan, JF Brady, Swim pressure: stress generation in active
  matter.
\newblock {\em\protect\JournalTitle{Physical review letters}} \textbf{113},
  028103 (2014).

\bibitem{omar2020microscopic}
AK Omar, ZG Wang, JF Brady, Microscopic origins of the swim pressure and the
  anomalous surface tension of active matter.
\newblock {\em\protect\JournalTitle{Physical Review E}} \textbf{101}, 012604
  (2020).

\bibitem{angelani2009self}
L Angelani, R Di~Leonardo, G Ruocco, Self-starting micromotors in a bacterial
  bath.
\newblock {\em\protect\JournalTitle{Physical review letters}} \textbf{102},
  048104 (2009).

\bibitem{reichhardt2017ratchet}
CO Reichhardt, C Reichhardt, Ratchet effects in active matter systems.
\newblock {\em\protect\JournalTitle{Annual Review of Condensed Matter Physics}}
  \textbf{8}, 51--75 (2017).

\bibitem{ro2022model}
S Ro, et~al., Model-free measurement of local entropy production and
  extractable work in active matter.
\newblock {\em\protect\JournalTitle{Physical Review Letters}} \textbf{129},
  220601 (2022).

\bibitem{damasceno2012predictive}
PF Damasceno, M Engel, SC Glotzer, Predictive self-assembly of polyhedra into
  complex structures.
\newblock {\em\protect\JournalTitle{Science}} \textbf{337}, 453--457 (2012).

\bibitem{wang2012colloids}
Y Wang, et~al., Colloids with valence and specific directional bonding.
\newblock {\em\protect\JournalTitle{Nature}} \textbf{491}, 51--55 (2012).

\bibitem{vlasov2001chip}
YA Vlasov, XZ Bo, JC Sturm, DJ Norris, On-chip natural assembly of silicon
  photonic bandgap crystals.
\newblock {\em\protect\JournalTitle{Nature}} \textbf{414}, 289--293 (2001).

\bibitem{liu2014orientationally}
Q Liu, MG Campbell, JS Evans, II Smalyukh, Orientationally ordered colloidal
  co-dispersions of gold nanorods and cellulose nanocrystals.
\newblock {\em\protect\JournalTitle{Advanced Materials}} \textbf{26},
  7178--7184 (2014).

\bibitem{engel2015computational}
M Engel, PF Damasceno, CL Phillips, SC Glotzer, Computational self-assembly of
  a one-component icosahedral quasicrystal.
\newblock {\em\protect\JournalTitle{Nature materials}} \textbf{14}, 109--116
  (2015).

\bibitem{wang2017colloidal}
Y Wang, IC Jenkins, JT McGinley, T Sinno, JC Crocker, Colloidal crystals with
  diamond symmetry at optical lengthscales.
\newblock {\em\protect\JournalTitle{Nature communications}} \textbf{8}, 14173
  (2017).

\bibitem{he2020colloidal}
M He, et~al., Colloidal diamond.
\newblock {\em\protect\JournalTitle{Nature}} \textbf{585}, 524--529 (2020).

\bibitem{ni2014crystallizing}
R Ni, MAC Stuart, M Dijkstra, PG Bolhuis, Crystallizing hard-sphere glasses by
  doping with active particles.
\newblock {\em\protect\JournalTitle{Soft Matter}} \textbf{10}, 6609--6613
  (2014).

\bibitem{mallory2020universal}
S Mallory, M Bowers, A Cacciuto, Universal reshaping of arrested colloidal gels
  via active doping.
\newblock {\em\protect\JournalTitle{The Journal of Chemical Physics}}
  \textbf{153}, 084901 (2020).

\bibitem{ramananarivo2019activity}
S Ramananarivo, E Ducrot, J Palacci, Activity-controlled annealing of colloidal
  monolayers.
\newblock {\em\protect\JournalTitle{Nature communications}} \textbf{10}, 3380
  (2019).

\bibitem{saud2023microdynamics}
KT Saud, MJ Solomon, Microdynamics of active particles in defect-rich colloidal
  crystals.
\newblock {\em\protect\JournalTitle{Journal of Colloid and Interface Science}}
  (2023).

\bibitem{omar2018swimming}
AK Omar, Y Wu, ZG Wang, JF Brady, Swimming to stability: structural and
  dynamical control via active doping.
\newblock {\em\protect\JournalTitle{ACS nano}} \textbf{13}, 560--572 (2018).

\bibitem{sastry2001relationship}
S Sastry, The relationship between fragility, configurational entropy and the
  potential energy landscape of glass-forming liquids.
\newblock {\em\protect\JournalTitle{Nature}} \textbf{409}, 164--167 (2001).

\bibitem{sastry1998signatures}
S Sastry, PG Debenedetti, FH Stillinger, Signatures of distinct dynamical
  regimes in the energy landscape of a glass-forming liquid.
\newblock {\em\protect\JournalTitle{Nature}} \textbf{393}, 554--557 (1998).

\bibitem{heuer2008exploring}
A Heuer, Exploring the potential energy landscape of glass-forming systems:
  from inherent structures via metabasins to macroscopic transport.
\newblock {\em\protect\JournalTitle{Journal of Physics: Condensed Matter}}
  \textbf{20}, 373101 (2008).

\bibitem{berthier2011dynamic}
L Berthier, Dynamic heterogeneity in amorphous materials.
\newblock {\em\protect\JournalTitle{arXiv preprint arXiv:1106.1739}} (2011).

\bibitem{berthier2011overview}
L Berthier, G Biroli, JP Bouchaud, RL Jack, Overview of different
  characterisations of dynamic heterogeneity.
\newblock {\em\protect\JournalTitle{Dynamical heterogeneities in glasses,
  colloids, and granular media}} \textbf{150}, 68 (2011).

\bibitem{wittmer2013shear}
J Wittmer, H Xu, P Poli{\'n}ska, F Weysser, J Baschnagel, Shear modulus of
  simulated glass-forming model systems: Effects of boundary condition,
  temperature, and sampling time.
\newblock {\em\protect\JournalTitle{The Journal of chemical physics}}
  \textbf{138}, 12A533 (2013).

\bibitem{yoshino2014shear}
H Yoshino, F Zamponi, Shear modulus of glasses: Results from the full
  replica-symmetry-breaking solution.
\newblock {\em\protect\JournalTitle{Physical Review E}} \textbf{90}, 022302
  (2014).

\bibitem{goodrich2016scaling}
CP Goodrich, AJ Liu, JP Sethna, Scaling ansatz for the jamming transition.
\newblock {\em\protect\JournalTitle{Proceedings of the National Academy of
  Sciences}} \textbf{113}, 9745--9750 (2016).

\bibitem{cao2019potential}
P Cao, MP Short, S Yip, Potential energy landscape activations governing
  plastic flows in glass rheology.
\newblock {\em\protect\JournalTitle{Proceedings of the National Academy of
  Sciences}} \textbf{116}, 18790--18797 (2019).

\bibitem{masoero2012nanostructure}
E Masoero, E Del~Gado, RM Pellenq, FJ Ulm, S Yip, Nanostructure and
  nanomechanics of cement: polydisperse colloidal packing.
\newblock {\em\protect\JournalTitle{Physical review letters}} \textbf{109},
  155503 (2012).

\bibitem{ioannidou2016mesoscale}
K Ioannidou, et~al., Mesoscale texture of cement hydrates.
\newblock {\em\protect\JournalTitle{Proceedings of the National Academy of
  Sciences}} \textbf{113}, 2029--2034 (2016).

\bibitem{zhou2019multiscale}
T Zhou, K Ioannidou, FJ Ulm, MZ Bazant, RM Pellenq, Multiscale poromechanics of
  wet cement paste.
\newblock {\em\protect\JournalTitle{Proceedings of the National Academy of
  Sciences}} \textbf{116}, 10652--10657 (2019).

\bibitem{zhou2020freezing}
T Zhou, M Mirzadeh, RJM Pellenq, MZ Bazant, Freezing point depression and
  freeze-thaw damage by nanofluidic salt trapping.
\newblock {\em\protect\JournalTitle{Physical Review Fluids}} \textbf{5}, 124201
  (2020).

\bibitem{monfared2020effect}
S Monfared, et~al., Effect of confinement on capillary phase transition in
  granular aggregates.
\newblock {\em\protect\JournalTitle{Physical Review Letters}} \textbf{125},
  255501 (2020).

\bibitem{szakasits2019rheological}
ME Szakasits, KT Saud, X Mao, MJ Solomon, Rheological implications of embedded
  active matter in colloidal gels.
\newblock {\em\protect\JournalTitle{Soft Matter}} \textbf{15}, 8012--8021
  (2019).

\bibitem{saud2021yield}
KT Saud, M Ganesan, MJ Solomon, Yield stress behavior of colloidal gels with
  embedded active particles.
\newblock {\em\protect\JournalTitle{Journal of Rheology}} \textbf{65}, 225--239
  (2021).

\bibitem{wei2023reconfiguration}
M Wei, MB Zion, O Dauchot, Reconfiguration, interrupted aging and enhanced
  dynamics of a colloidal gel using photo-switchable active doping.
\newblock {\em\protect\JournalTitle{arXiv preprint arXiv:2302.08360}} (2023).

\bibitem{hueckel2020ionic}
T Hueckel, GM Hocky, J Palacci, S Sacanna, Ionic solids from common colloids.
\newblock {\em\protect\JournalTitle{Nature}} \textbf{580}, 487--490 (2020).

\bibitem{grigera2001fast}
TS Grigera, G Parisi, Fast monte carlo algorithm for supercooled soft spheres.
\newblock {\em\protect\JournalTitle{Physical Review E}} \textbf{63}, 045102
  (2001).

\bibitem{ninarello2017models}
A Ninarello, L Berthier, D Coslovich, Models and algorithms for the next
  generation of glass transition studies.
\newblock {\em\protect\JournalTitle{Physical Review X}} \textbf{7}, 021039
  (2017).

\bibitem{ozawa2018random}
M Ozawa, L Berthier, G Biroli, A Rosso, G Tarjus, Random critical point
  separates brittle and ductile yielding transitions in amorphous materials.
\newblock {\em\protect\JournalTitle{Proceedings of the National Academy of
  Sciences}} \textbf{115}, 6656--6661 (2018).

\bibitem{wang2015constant}
M Wang, JF Brady, Constant stress and pressure rheology of colloidal
  suspensions.
\newblock {\em\protect\JournalTitle{Physical review letters}} \textbf{115},
  158301 (2015).

\bibitem{alexander1998amorphous}
S Alexander, Amorphous solids: their structure, lattice dynamics and
  elasticity.
\newblock {\em\protect\JournalTitle{Physics reports}} \textbf{296}, 65--236
  (1998).

\bibitem{zaccone2009elasticity}
A Zaccone, H Wu, E Del~Gado, Elasticity of arrested short-ranged attractive
  colloids: Homogeneous and heterogeneous glasses.
\newblock {\em\protect\JournalTitle{Physical review letters}} \textbf{103},
  208301 (2009).

\bibitem{maxwell1864calculation}
JC Maxwell, L. on the calculation of the equilibrium and stiffness of frames.
\newblock {\em\protect\JournalTitle{The London, Edinburgh, and Dublin
  Philosophical Magazine and Journal of Science}} \textbf{27}, 294--299 (1864).

\bibitem{whitaker2019colloidal}
KA Whitaker, et~al., Colloidal gel elasticity arises from the packing of
  locally glassy clusters.
\newblock {\em\protect\JournalTitle{Nature communications}} \textbf{10}, 2237
  (2019).

\bibitem{hsiao2012role}
LC Hsiao, RS Newman, SC Glotzer, MJ Solomon, Role of isostaticity and
  load-bearing microstructure in the elasticity of yielded colloidal gels.
\newblock {\em\protect\JournalTitle{Proceedings of the National Academy of
  Sciences}} \textbf{109}, 16029--16034 (2012).

\bibitem{singh2020brittle}
M Singh, M Ozawa, L Berthier, Brittle yielding of amorphous solids at finite
  shear rates.
\newblock {\em\protect\JournalTitle{Physical Review Materials}} \textbf{4},
  025603 (2020).

\bibitem{pollard2022yielding}
J Pollard, SM Fielding, Yielding, shear banding, and brittle failure of
  amorphous materials.
\newblock {\em\protect\JournalTitle{Physical Review Research}} \textbf{4},
  043037 (2022).

\bibitem{rossi2022finite}
S Rossi, G Biroli, M Ozawa, G Tarjus, F Zamponi, Finite-disorder critical point
  in the yielding transition of elastoplastic models.
\newblock {\em\protect\JournalTitle{Physical Review Letters}} \textbf{129},
  228002 (2022).

\bibitem{rosenfeld1977relation}
Y Rosenfeld, Relation between the transport coefficients and the internal
  entropy of simple systems.
\newblock {\em\protect\JournalTitle{Physical Review A}} \textbf{15}, 2545
  (1977).

\bibitem{dyre2018perspective}
JC Dyre, Perspective: Excess-entropy scaling.
\newblock {\em\protect\JournalTitle{The Journal of chemical physics}}
  \textbf{149}, 210901 (2018).

\bibitem{tang2021energy}
L Tang, et~al., The energy landscape governs ductility in disordered materials.
\newblock {\em\protect\JournalTitle{Materials Horizons}} \textbf{8}, 1242--1252
  (2021).

\bibitem{kim2022active}
Y Kim, S Joo, WK Kim, JH Jeon, Active diffusion of self-propelled particles in
  flexible polymer networks.
\newblock {\em\protect\JournalTitle{Macromolecules}} \textbf{55}, 7136--7147
  (2022).

\bibitem{yadav2023dynamics}
RS Yadav, C Das, R Chakrabarti, Dynamics of a spherical self-propelled tracer
  in a polymeric medium: interplay of self-propulsion, stickiness, and
  crowding.
\newblock {\em\protect\JournalTitle{Soft Matter}} \textbf{19}, 689--700 (2023).

\bibitem{kumar2023dynamics}
P Kumar, R Chakrabarti, Dynamics of self-propelled tracer particles inside a
  polymer network.
\newblock {\em\protect\JournalTitle{Physical Chemistry Chemical Physics}}
  \textbf{25}, 1937--1946 (2023).

\bibitem{ben2015modeling}
E Ben-Isaac, {\'E} Fodor, P Visco, F Van~Wijland, NS Gov, Modeling the dynamics
  of a tracer particle in an elastic active gel.
\newblock {\em\protect\JournalTitle{Physical Review E}} \textbf{92}, 012716
  (2015).

\bibitem{paul2023dynamical}
K Paul, A Mutneja, SK Nandi, S Karmakar, Dynamical heterogeneity in active
  glasses is inherently different from its equilibrium behavior.
\newblock {\em\protect\JournalTitle{Proceedings of the National Academy of
  Sciences}} \textbf{120}, e2217073120 (2023).

\bibitem{mallory2019activity}
SA Mallory, A Cacciuto, Activity-enhanced self-assembly of a colloidal kagome
  lattice.
\newblock {\em\protect\JournalTitle{Journal of the American Chemical Society}}
  \textbf{141}, 2500--2507 (2019).

\bibitem{pinson2018inferring}
MB Pinson, T Zhou, HM Jennings, MZ Bazant, Inferring pore connectivity from
  sorption hysteresis in multiscale porous media.
\newblock {\em\protect\JournalTitle{Journal of colloid and interface science}}
  \textbf{532}, 118--127 (2018).

\bibitem{simkiss2012biomineralization}
K Simkiss, KM Wilbur, {\em Biomineralization}.
\newblock (Elsevier), (2012).

\bibitem{nodehi2022systematic}
M Nodehi, T Ozbakkaloglu, A Gholampour, A systematic review of bacteria-based
  self-healing concrete: Biomineralization, mechanical, and durability
  properties.
\newblock {\em\protect\JournalTitle{Journal of Building Engineering}}
  \textbf{49}, 104038 (2022).

\bibitem{bullard2011mechanisms}
JW Bullard, et~al., Mechanisms of cement hydration.
\newblock {\em\protect\JournalTitle{Cement and concrete research}} \textbf{41},
  1208--1223 (2011).

\bibitem{wang2015synthetic}
Y Wang, et~al., Synthetic strategies toward dna-coated colloids that
  crystallize.
\newblock {\em\protect\JournalTitle{Journal of the American Chemical Society}}
  \textbf{137}, 10760--10766 (2015).

\bibitem{stukowski2014computational}
A Stukowski, Computational analysis methods in atomistic modeling of crystals.
\newblock {\em\protect\JournalTitle{Jom}} \textbf{66}, 399--407 (2014).

\bibitem{ackland2006applications}
G Ackland, A Jones, Applications of local crystal structure measures in
  experiment and simulation.
\newblock {\em\protect\JournalTitle{Physical Review B}} \textbf{73}, 054104
  (2006).

\end{thebibliography}

\end{document}